\documentclass[preprint,aps,prl,floatfix,a4paper,onecolumn]{revtex4}
\usepackage{color}
\usepackage[english]{babel}
\makeatletter\AtBeginDocument{\let\@elt\relax}\makeatother
\usepackage{graphicx}
\usepackage{dcolumn}
\usepackage{amsmath}

\usepackage{setspace}
\usepackage[justification=justified]{caption}
\def\beq#1{\begin{equation}\label{#1}}
\def\eeq{\end{equation}}

\begin{document}
\title{Enhanced chiral-sensitivity of Coulomb-focused electrons in strong field ionization}

\author{
S. Rozen$^{1}$, 
S. Larroque$^{2}$, 
N. Dudovich$^{1}$, 
Y. Mairesse$^{2}$,
B. Pons$^{2}$
}
\bigskip

\affiliation{
	$^1$ Weizmann Institute of Science, Rehovot, 76100, Israel\\
	$^2$ Universit\'e de Bordeaux - CNRS - CEA, CELIA, UMR5107, F33405 Talence, France
}

\date{\today}

\begin{abstract}
Strong-field light-matter interactions initiate a wide range of phenomena in which the quantum paths of 
electronic wavepackets can be manipulated by tailoring the laser field. 
Among the electrons released by a strong laser pulse from atomic and molecular targets, 
some are subsequently driven back to the vicinity of the ionic core by the oscillating laser field.
The trajectories of these returning electrons are bent towards the core by the ionic potential,
an effect known as Coulomb focusing. This process, studied over the past two decades, has been 
associated with the long range influence of the Coulomb potential. 
Here we explore the structural properties of the Coulomb focusing phenomenon. Specifically, we numerically 
study the sensitivity
of the returning electron dynamics to the anisotropy of the ionic potential. We employ orthogonally polarized
two-color strong fields and chiral molecules, whose asymmetric features lead to unambiguous fingerprints
of the potential on the freed electrons. The Coulomb-focused electrons show an enhanced sensitivity to
chirality, related to an asymmetric attoclock-like angular streaking stemming from field-assisted scattering of the
electrons onto the chiral ionic potential. Anisotropic features of the ionic potential thus
monitor the motion of Coulomb-focused electrons throughout their returning paths, 
shedding light on the structural properties of the interaction.
\end{abstract}
\maketitle
	
\section{Introduction}

The interaction of strong laser fields with atoms or molecules triggers coherent electronic 
dynamics on the attosecond timescale, which can be used to probe the structure and 
dynamics of the irradiated target with 
unique spatial and temporal resolutions. Electron wavepackets are released by tunnel-ionization around the maxima of 
the laser field oscillation, and are subsequently accelerated by the laser field. The electrons born
before the field maxima directly escape the ionic potential to form above-threshold ionization (ATI) peaks 
\cite{agostini1979}. These electrons are identified as "direct" electrons. 
By contrast, the electrons born after the field maxima are driven back to the vicinity of their parent ion, 
where they can be subject 
to: 
(i) elastic forward scattering onto the ionic core, yielding the "indirect" contribution to the ATI 
distribution;
(ii) elastic backward scattering, leading to high order ATI and laser-induced electron diffraction \cite{paulus1994,meckel2008,blaga2012}; 
(ii) inelastic scattering, producing non-sequential double ionization \cite{lhuillier1983};
(iii) radiative recombination with the ion to produce bursts of attosecond extreme ultraviolet radiation, 
forming high order harmonics \cite{mcpherson1987,ferray1988,paul2001a}. 

Stating the important role of the ionic potential in the scattering process seems like a truism. Nevertheless some of its aspects are 
often neglected in the description and interpretation of the strong-field spectroscopic techniques listed 
above. Indeed, the strong-laser field generally dominates the interaction, such that the Strong-Field Approximation (SFA) is an excellent approach to describe many properties of the electron dynamics \cite{lewenstein1994,amini2019}. One important correction to this description is Coulomb focusing \cite{brabec1996}. The Coulomb force exerted by the ionic potential bends the returning electron trajectories towards 
the ionic core, focusing the wavepacket. This has important consequences on 
high-harmonic generation (HHG, \cite{shafir2012}) and non-sequential double ionization \cite{brabec1996}. 
However the influence of Coulomb focusing on the indirect electrons in ATI is more elusive. It  
has been shown that the focusing reduces the transverse momentum distribution of the ionized electrons 
\cite{comtois2005,shafir2013,li2013,landsman2013a}. Recently, a spectacular signature of Coulomb focusing has been observed 
in strong-field ionization of argon atoms by orthogonally polarized two-color (OTC) laser fields \cite{richter2016}. 
In this configuration, a weak second-harmonic field is added to the fundamental one to manipulate the 
electron trajectories. By measuring the photoelectron angular distribution in the laser polarization plane, 
Richter \textit{et al.} found out that some of the electrons remained rather insensitive to the second-harmonic field, 
seemingly ignoring the lateral momentum streaking resulting from the trajectory manipulation. 
These "unstreakable" electrons were identified as being indirect electrons that have experienced Coulomb focusing. 
The OTC field configuration is thus particularly adequate to reveal the underlying spatial and temporal properties 
of the Coulomb focusing. 

In the framework of HHG, Coulomb focusing was found to influence 
mainly the long trajectories \cite{shafir2012}. Therefore it was interpreted in terms of a long-range effect, 
dictated by the asymptotic, Coulombic, part of the ionic potential. The influence of the detailed target structure and 
associated potential geometry on Coulomb focusing has thus never been considered so far.  

In this paper, we numerically investigate the sensitivity of Coulomb-focused electrons to the 
structure of the ionized target. 
To emphasize the influence of the ionic potential, we use a target with a specific geometrical property: 
chiral molecules, which are not superimposable to their mirror image. When chiral molecules are photoionized 
by circularly polarized light, the photoelectron momentum distribution is forward/backward asymmetric with 
respect to the light propagation axis. This effect, called PhotoElectron Circular Dichroism (PECD), produces 
large chirosensitive signals, resulting from pure electric-dipole interaction. 
PECD was initially predicted \cite{ritchieTheoryAngularDistribution1976,powis2000} and measured \cite{bowering2001,janssen2013,nahon2015} in single-photon ionization by extreme ultraviolet 
radiation. It was interpreted as resulting from the influence of the chiral potential on the scattering of the outgoing electrons in the rotating electromagnetic field. Classically, this asymmetry can be seen as analogue to the conversion of the rotation motion of a nut into directional translation on a bolt \cite{powis2008a}. Quantum mechanically, the PECD is dictated by the phase shifts of the continuum scattering wavefunctions. For more information we refer the reader to the tutorial paper by Powis \cite{powis2008a}.  PECD was later extended to the multiphoton \cite{lux2012,lehmann2013} and strong-field \cite{beaulieu2016} 
regimes. Strong-field PECD combines the advantages of strong-field spectroscopy -- orbital selectivity, attosecond 
temporal resolution and Angstr\"om spatial resolution -- with the unambiguous structural sensitivity of chiroptical measurements. 
It was used to measure the attosecond delays between ionization of a molecule and its mirror image 
\cite{beaulieu2017a}, to relate PECD to different fragmentation pathways associated with different molecular 
alignments \cite{fehre2019a}, to investigate the chirosensitivity of angular streaking \cite{fehre2019b,bloch2021}, 
and to probe the influence of the instantaneous chirality of light to chiral light-matter interaction \cite{rozen2019}. 

Here, strong-field PECD is used to reveal the chiral sensitivity of Coulomb-focused 
electrons. We employ a toy-model chiral molecule which was shown to produce chiroptical signals similar to genuine
chiral molecules \cite{rozen2019}. The chiral interaction is driven by OTC fields which yield an unambiguous 
fingerprint of indirect electrons, Coulomb-focused along the fundamental component of the field \cite{richter2016}. 
These electrons exhibit enhanced sensitivity to chirality. We show that molecular chirality is encoded in 
the Coulomb-focused signal in terms of a forward/backward asymmetric angular streaking of the electrons, due 
to laser-assisted scattering of the electrons onto the chiral ionic potential.

\section{Identifying Coulomb focusing in photoelectron angular distributions}
PECD measurements are conventionally performed using circularly polarized radiation. 
In the strong-field regime, the rotating laser field lowers the potential barrier of the molecule, 
continuously releasing electrons by tunnel ionization. These electrons are accelerated by the field and 
directly escape the molecular potential, without coming back to the vicinity of the ion. 
The forward/backward asymmetry in the electron angular distribution arises from both the electron scattering off 
the ionic potential \cite{powis2008a} and the sensitivity of tunneling dynamics to the 
chirality of the barrier \cite{bloch2021}. As soon as the polarization state is not circular, the magnitude of 
the electric field oscillates in time and tunneling releases multiple families of electron trajectories in 
the continuum. Manipulating the vectorial properties of the laser field oscillations provides a unique way 
to finely control the chiroptical interaction \cite{rozen2019}, by shaping the electron trajectories 
\cite{kitzler2005} as well as the time-dependent optical chirality \cite{neufeld2018} of the ionizing radiation. 

In this work, we use an OTC field ${\bf E}$ defined as:
\begin{equation}
	{\bf E}(t)=E_{0}\cos\left(\omega t\right){\bf \hat{y}}+E_{0} \sqrt{r}\cos\left(2\omega t+\varphi_{\omega/2\omega}\right){\bf \hat{x}}
\end{equation}
where $E_0$ is the amplitude of the fundamental field with frequency $\omega$, $r=I_{2\omega}/I_\omega < 1$ 
is the ratio between the second harmonic and fundamental field intensities, 
and $\varphi_{\omega/2\omega}$ is the relative phase between 
the two frequency components. Varying $\varphi_{\omega/2\omega}$ allows to tune the OTC field shape which
switches from an "8", when $\varphi_{\omega/2\omega}=\pi/2\pm\pi$, to a "C" when $\varphi_{\omega/2\omega}=0\pm \pi$.  

The strong-field ionization dynamics in an OTC fields are very close to those in linearly polarized fields. 
The electrons emitted before the maxima of the laser field oscillation directly escape the molecular potential -- these
are the direct electrons. By contrast, the electrons ionized after the maxima of the field escape in one 
direction before being driven back to the vicinity of the ion when the field reverses -- they form the indirect 
electron family. The two electron families naturally show up in the SFA where the photoionization process is described 
in terms of quantum paths exclusively monitored by the strong driving field \cite{lewenstein1994,salieres2001}.
The phase of each quantum path is given by the action integral $S$ accumulated along the electron trajectory.
For a given final electron momentum ${\bf p}$, minimizing the action integral provides 
the dominant quantum paths $j$, with $j=1$ and 2 for direct and indirect electrons, respectively. The photoelectron momentum 
distribution is thus $\mathcal{P}({\bf p}) = \left| \sum_{j} e^{i\,S({\bf p},t_0^j)}\right|^2$, where $t_0^j$ is the ionization 
time of electron $j$. The influence of the ionic potential on the escaping electrons is neglected in the SFA framework. 
As a result, the final momentum ${\bf p}$ of the electrons is completely determined by the value of the vector potential 
${\bf A}$ at the time of ionization $t_0$: ${\bf p}=-{\bf A}(t_0)$. Direct and indirect electron paths thus asymptotically lead 
to identical momentum distributions when $\varphi_{\omega/2\omega}=\pi/2$. This is illustrated in the inset of Fig. 1(a) for 
ionization of a target with ionization potential $I_P=7.2$ eV by an OTC field with $I_\omega= 5\times 10^{13}$ 
W.cm$^{-2}$ and $r=0.1$. The shape of direct and indirect momentum distributions indeed 
conforms to the "C"-shape of $-{\bf A}$, and their overlap results in interference patterns along the 
multiple concentric shells associated to the ATI peaks (see Fig. 1(a)). When the relative phase 
$\varphi_{\omega/2\omega}$ is switched to 0, the direct and indirect electrons are streaked towards opposite directions
along the $x$-axis since the vector potential exhibits an "8"-shape in that case. However, the direct and indirect 
asymptotic wavepackets still significantly overlap, and produce clear interference patterns along the ATI rings, 
as shown in Fig. 1(d).

\begin{figure}[t]\centering 
	\includegraphics[width=0.8\linewidth]{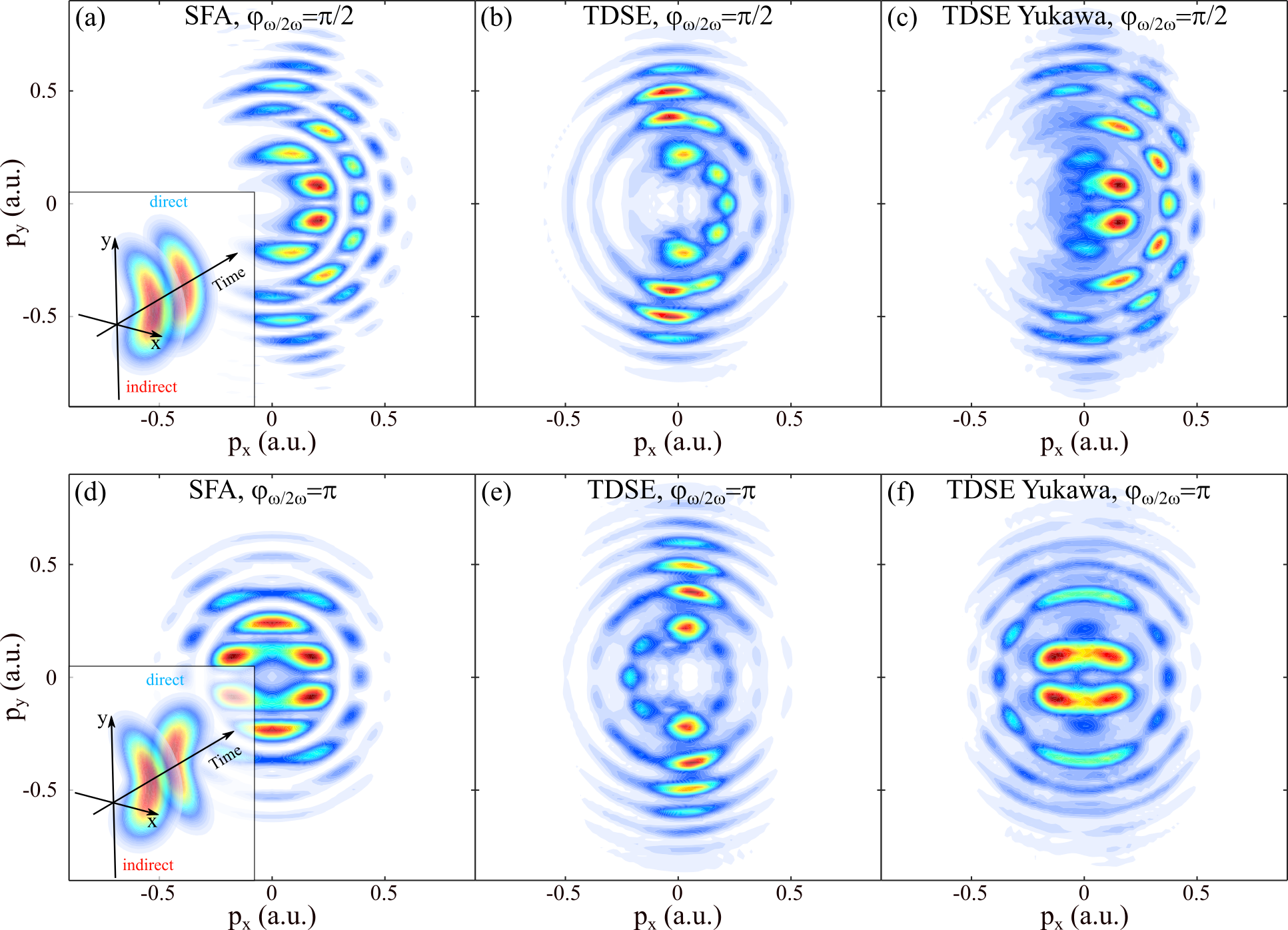}
\caption{(a) Photoelectron angular distribution (PAD) in the polarization plane of a 800nm/400nm OTC field
with $I_\omega=5\times 10^{13}$ W.cm$^{-2}$, $r=0.1$ and $\varphi_{\omega/2\omega}\pi/2$, obtained by SFA
when ionizing a target with ionization potential $I_P=7.2$ eV. The concentric radial ATI shells are angularly
structured because of the interference of direct and indirect wavepackets schematically represented
in the inset.
(b) Result of TDSE calculations for ionization of the toy-model chiral molecule by the same OTC field.
(c) Result of TDSE calculations for ionization of the long-range Yukawa-screened molecule.
(d-f) Same as (a-c) but for an OTC field with $\varphi_{\omega/2\omega}=\pi$. }
	\label{fig1}
\end{figure}

The SFA calculations provide an intuitive picture of the interaction, but they neglect the influence of the ionic 
potential on the ionization dynamics. They are thus unable not only to describe Coulomb focusing but also to produce any PECD signal, 
as demonstrated in \cite{dreissigacker2014}. To investigate the impact of the potential, we solved the Time-Dependent 
Schr\"odinger Equation (TDSE) for the toy-model chiral system introduced in \cite{rozen2019,bloch2021}. 
The calculations strictly followed the numerical recipes described in \cite{rozen2019,bloch2021}. Briefly, the 
toy-model system consists of a single-electron evolving in the (chiral) field of four fictitious nuclei. Its ionization
potential is $I_P=9$ eV. We mimic 
the random orientation of the molecular sample by solving the TDSE for many molecular orientations defined in the 
laboratory frame. The TDSE is solved in the velocity gauge by expanding the total electron wavefunction 
onto an underlying set of spherical Bessel functions. The 
photoelectron momentum distribution is computed at the end of the interaction as the sum of the 
ionizing densities associated to all molecular orientations. The driving OTC field is a flat-top four-cycle pulse
with 800 nm fundamental wavelength, including one $\omega$-cycle ascending and descending ramps at the beginning 
and at the end of the interaction, respectively. The photoelectron momentum distributions $P({\bf p})$ obtained 
in the polarization plane of OTC fields with $I_\omega= 5\times 10^{13}$ W.cm$^{-2}$, $r=0.1$, 
$\varphi_{\omega/2\omega}=\pi/2$ and 0 are shown in Fig. 1(b) and Fig. 1(e), respectively. The results are 
strikingly different from the SFA calculations. The TDSE distributions are dominated by a sharp vertical component, 
localized around the polarization direction of the fundamental laser field. Furthermore, this component remains 
largely unaffected by the relative phase between the fundamental and second harmonic fields. Such a feature 
was observed in the photoionization of argon atoms by Richter \textit{et al.}, who labeled this component 
"unstreakable", because of its insensitivity to the second harmonic field \cite{richter2016}. By using classical 
calculations, they showed that this component resulted from the Coulomb focusing of indirect electrons. 

In order to confirm the origin of the sharp component in the momentum distribution, we damped the long-range 
part of the potential of our target molecule and repeated the TDSE calculations. The screening was introduced 
by multiplying the molecular potential by an isotropic cut-off Yukawa term $\exp^{-(r-r_0)}$ for
radial distances $r$ larger than $r_0=3.5$ a.u., where $r_0$ represents the spatial extent of our molecular system. 
Since the ionic potential remains unchanged in the $r < r_0$ inner range, the Yukawa term makes the ionization 
potential of the screened system slightly smaller than its unscreened counterpart ($I_P=7.2$ eV). 
The screened potential vanishes for $r \gtrsim 8$ a.u.. The screening procedure thus cancels the 
influence of the long-range potential on electron scattering, as recently established by Torlina \textit{et al.} 
in atomic attoclock simulations \cite{torlina2015}. The momentum distributions from the screened molecules are 
presented in Fig. 1(c) and Fig. 1(f) for $\varphi_{\omega/2\omega}=\pi/2$ and 0, respectively. 
They are remarkably similar to the SFA distributions of Fig. 1(a) and Fig. 1(d), confirming that the localized
vertical component observed in the unscreened calculations results from the influence of the long-range potential 
in the ionization dynamics. 

The fact that Coulomb focusing disappears when the ionic potential is screened indicates that it is a long-range 
effect. This is consistent with the conclusions drawn from HHG driven by 
elliptical fields, where Coulomb focusing was found to influence mainly the long trajectories \cite{shafir2012}. 
Here, long range explicitly means $r \gtrsim 8$ a.u., which is the boundary beyond which the screened
potential vanishes. Coulomb focusing is thus dictated by this outer part of the potential which tends towards 
its isotropic Coulombic limit as $r \rightarrow \infty$. However, below this limit the unscreened potential is chiral, 
so that the focusing dynamics could present asymmetric features. 
Can we observe a chiral response of the Coulomb-focused electrons?

\section{Chiral sensitivity of Coulomb-focused electrons}
To reveal the influence of molecular chirality on photoionization, we rely on the rotation of the OTC field in time. 
This rotation reverses every half cycle of the fundamental field. Therefore, the OTC field thus carries 
zero net chirality. 
However its instantaneous chirality $C(t)$ can be defined as \cite{neufeld2018}:
\begin{equation}
C(t) = C_0 (E_x(t) \partial_t E_y(t) - E_y(t) \partial_t E_x(t))
\end{equation}
where $C_0$ is a normalization factor defined such that $C(t)=1$ for a right circularly polarized field.  
This quantity describes the rotation speed of the electric field. As shown theoretically in 
\cite{demekhin2018,rozen2019} and experimentally in \cite{rozen2019}, even if their instantaneous chirality reverses 
every half cycle, OTC fields can produce chiroptical signals in photoionization. This is 
due to the fact that 
the chiral response is accumulated while the departing electron is in the vicinity of the molecule, i.e. over a 
few hundreds of attoseconds during which the instantaneous chirality can keep a well defined sign. The variations 
of $C(t)$ induce a temporal gating of the chiral response, to which we referred to as ESCARGOT 
(Enantiosensitive SubCycle Antisymmetric Response Gated by electric-field rOTation) \cite{rozen2019}. 

\begin{figure}[t]\centering 
	\includegraphics[width=\linewidth]{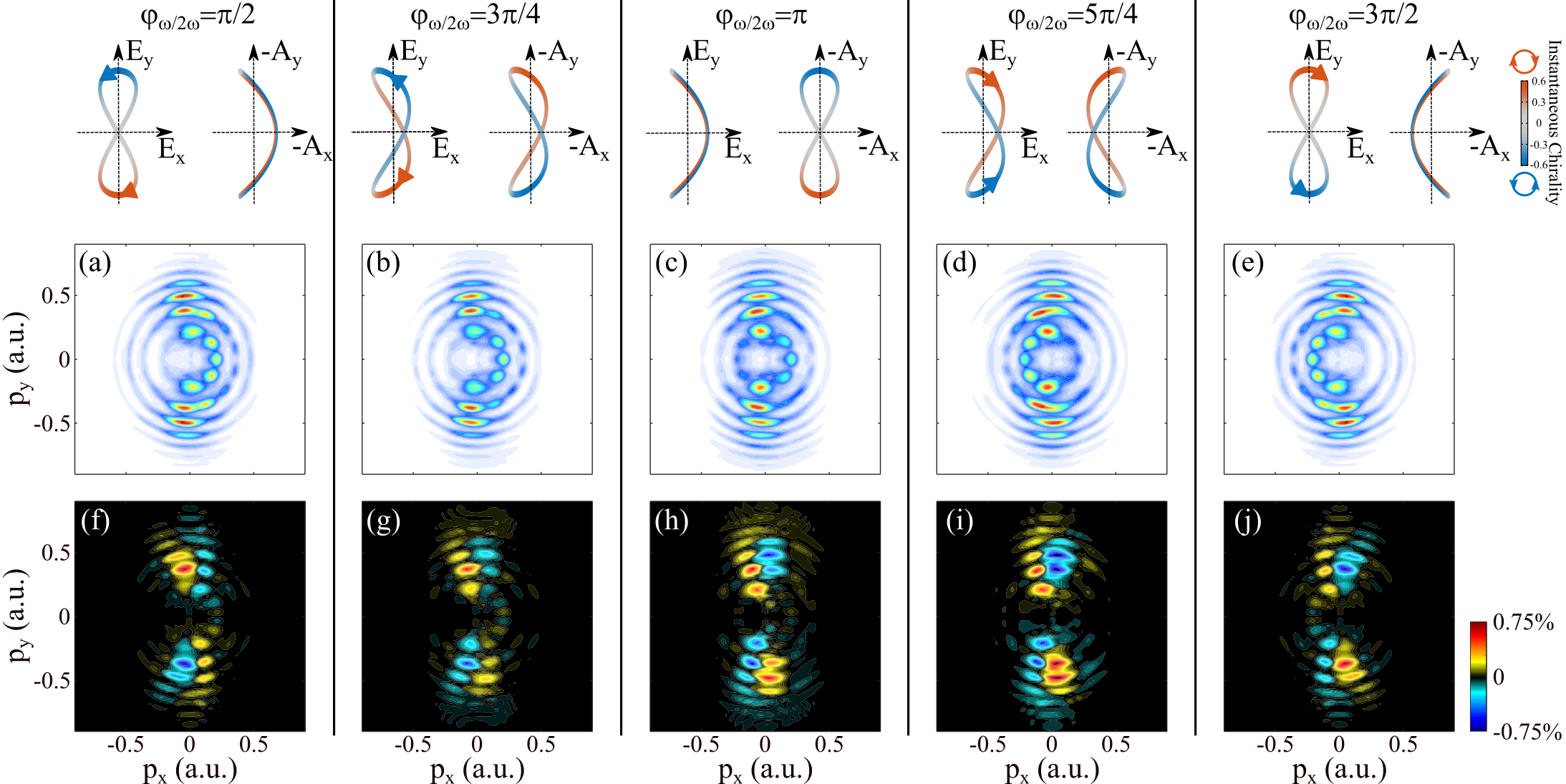}
	\caption{TDSE calculations of chiral photoionization by an OTC field. Top: shape of the electric field and opposite of the vector potential, as a function of the relative phase between the fundamental and second harmonic component of the OTC field. The color scale depicts the instantaneous optical chirality. (a-e) Cuts of the photoelectron angular distribution  in the $p_z=0.055$ a.u. plane, and (f-j) cuts of the forward backward asymmetry $\Delta P^{f/b}$, obtained by TDSE calculations in a toy-model chiral molecule ionized by  an OTC field with 5$\times 10^{13}$ W.cm$^{-2}$ and $r=0.1$.}
	\label{fig2}
\end{figure}

The use of OTC fields in chiral photoionization induces several characteristic symmetry properties.  First, 
the chiral response appears as a forward/backward asymmetry in the photoelectron angular distribution, as in 
conventional PECD. Secondly, as shown in Figure \ref{fig2}, the chirality of the field reverses every half cycle, 
for all two color phases. This leads to an up/down antisymmetry of the chiral response,
observed both in calculations 
and experiments \cite{demekhin2018,rozen2019}. 

To investigate the chirosensitive part of the signal, we extract the forward/backward antisymmetric component of
the photoelectron momentum distribution:
\begin{equation}
\Delta P^{f/b}(p_x,p_y,p_z)=[ P(p_x,p_y,p_z)-P(p_x,p_y,-p_z) ]/P_{max}
\end{equation}
that we normalize to the maximum of the 3D distribution, $P_{max}=\max(P({\bf p}))$.

The contribution of Coulomb-focused indirect electrons to $P({\bf p})$ maximizes about the $p_z=0$ polarization
plane \cite{comtois2005,shafir2013,li2013}. We thus present in Figure \ref{fig2} prototypical 
cuts of $\Delta P^{f/b}$ in the $p_z=0.055$ a.u. plane, for different values of $\varphi_{\omega/2\omega}$. 
For all shapes of the OTC field, the forward/backward asymmetry remains localized 
around the polarization direction ${\bf \hat{y}}$ of the fundamental laser field, in the area identified as originating 
from Coulomb-focused indirect trajectories. 
These results demonstrate the large influence of molecular chirality on Coulomb-focused electrons.  
A deeper insight into the 
process can be obtained by monitoring the evolution of the chiral signal as a function of 
$\varphi_{\omega/2\omega}$. In the upper hemisphere ($p_y>0$), the asymmetry is positive for all phases on the 
left edge ($p_x < 0$) of the momentum distribution and negative on the right edge. The situation is opposite 
in the lower hemisphere. Similarly to what we observed for the photoelectron momentum distributions, the 
whole chiral response of Coulomb-focused electrons is thus almost insensitive to the second harmonic component 
of the OTC field. This is a puzzling result since one would expect that electrons 
driven by an almost linearly polarized field will show a low chiral response. 
It is particularly remarkable that switching the relative phase from $\varphi_{\omega/2\omega}= \pi/2$ 
to $\varphi_{\omega/2\omega}= 3\pi/2$ reverses the helicity of the field but keeps the chiral signal positive 
in the upper left quadrant and negative in the upper right one. In the following we show that this can be 
intuitively understood as an effect of the angular streaking by the rotating laser field. 

\section{Angular streaking in orthogonal two-color laser fields}
\begin{figure}[t]\centering 
	\includegraphics[width=0.8\linewidth]{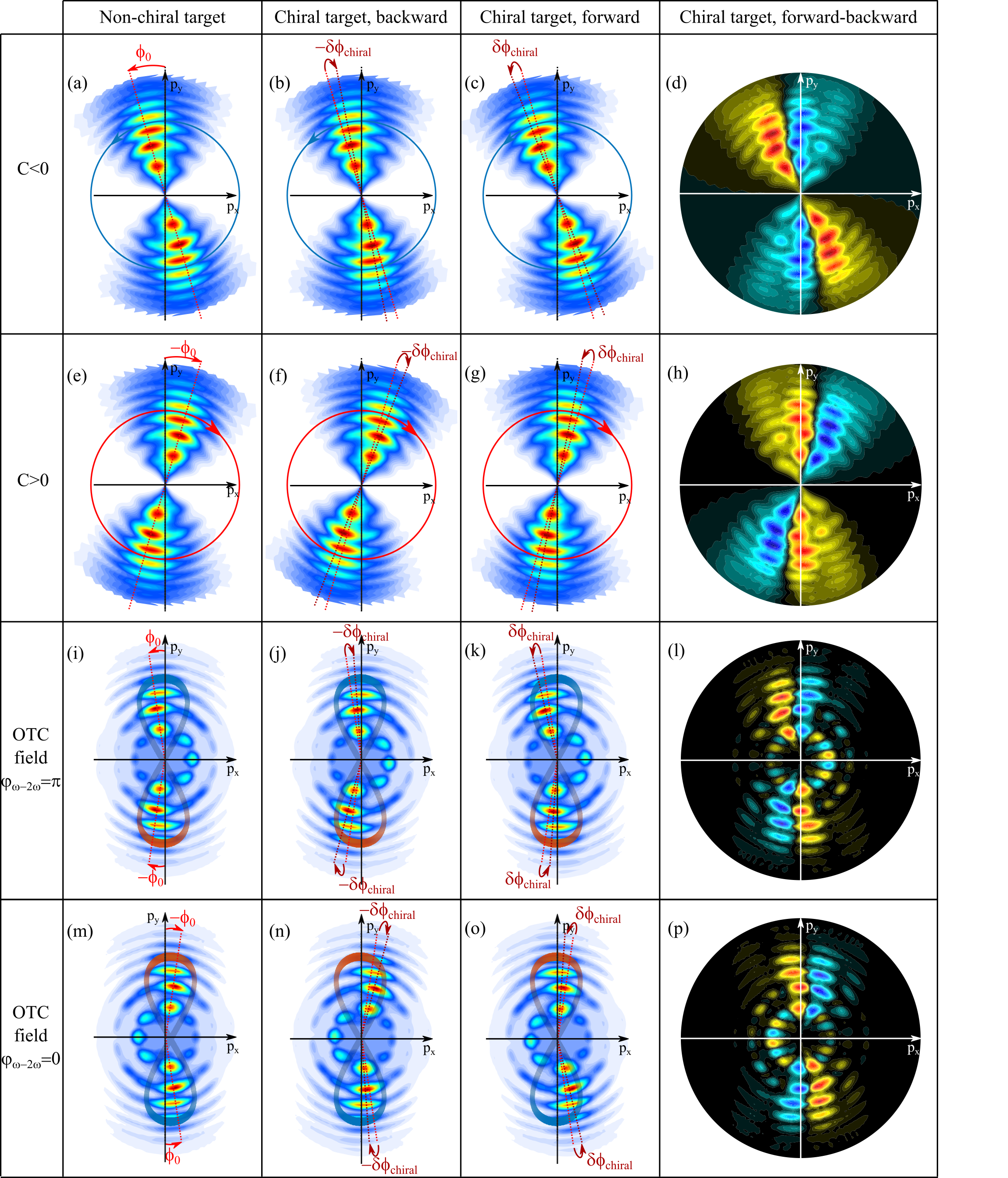}
\caption{Influence of angular streaking on the chiroptical response of Coulomb-focused electrons.
(a) Schematic representation of a photoelectron momentum distribution in the polarization plane of a
counter-clockwise rotating laser field. The electron distribution is shifted with respect to the
Coulomb-focusing $y$-direction by an angle $\phi_{0}$, which is determined by the scattering in
the potential.
(b,c) If the molecule is chiral, the distributions of electrons ejected in the backward (b) and forward (c)
hemispheres are shifted by the additional, and opposite, angles $\mp \delta\phi_{chiral}$.
(d) Asymmetric signal obtained by subtracting the forward (c) and backward (b) momentum distributions.
(e-h) Same schematic representation for a clockwise rotating field.
(i) TDSE calculations of the photoelectron momentum distribution in an OTC field with $I_{\omega}=5\times 10^{13}$ W.cm$^{-2}$, $r=0.1$,
$\varphi_{\omega/2\omega}=\pi$. The time-dependent rotation of
the vector potential, depicted by the color-coded line, induces opposite attoclock streaking angles
$\pm \phi_0$ in the upper and lower hemispheres, while molecular chirality leads to opposite
chiral streaking angles $\mp \delta\phi_{chiral}$ in the forward (j) and backward (k) hemispheres.
(l) Resulting forward/backward asymmetry.
(m-p) Same schematic representation than (i-l) but for a two-color phase $\delta\phi_{chiral}=0$.}
\label{figstreak}
\end{figure}
Angular streaking occurs when atoms or molecules are ionized by a rotating strong laser field, and has been 
extensively used to perform attoclock measurements \cite{eckle2008,landsman2014}. The angular streaking effect 
in a counter-clockwise rotating laser field is schematically depicted in Fig. \ref{figstreak}(a). The electrons 
tunnel out in a direction set by the laser field, and are angularly streaked by the rotation of the field. 
Within the strong field approximation, the final momentum of the electrons is dictated by the vector potential 
at the time of ionization. Deviations from this direction can be induced by delays in the tunneling process, or 
by scattering in the ionic potential at long range \cite{landsman2014,kheifets2020}. The latter effect generally 
dominates the interaction and induces an angular shift $\phi_{0}$ of the electron momentum distribution in the rotation 
direction of the field, as schematized in Fig. \ref{figstreak}(a) \cite{torlina2015,sainadh2019}. Reversing the 
helicity of the laser field switches the direction of the angular streaking, as shown in Fig \ref{figstreak}(e).

The phenomenon of angular streaking is not restricted to circular or elliptical polarization. Attoclock 
measurements can be performed with more complex laser fields, for instance combining counter-rotating or 
co-rotating fundamental and second harmonic beams \cite{eicke2019,bloch2021}. Angular streaking should thus 
play a role in the photoionization by OTC fields. The most intuitive situation in terms of angular streaking 
is the one where the relative phase between the two components of the OTC field is 
$\varphi_{\omega/2\omega}=\pi$ (Fig. \ref{figstreak}(i)). The vector potential describes an 8 shape. In the upper 
hemisphere, the potential rotates counter-clockwise. In the absence of interaction with the ionic potential, 
the electron distribution would maximize along $p_x=0$ according to SFA. The Coulomb-focused signal is clearly 
shifted to the left, i.e. towards positive attoclock streaking angles. The attoclock offset angle $\phi_{0}$ 
decreases from lower to higher ATI peaks. This indicates a lower influence of the potential on high kinetic 
energy electrons. In the lower hemisphere, the potential rotates clockwise. The momentum distribution is again 
shifted to the left, which also corresponds to a positive streaking angle with respect to the field rotation 
direction, but to an opposite shift $-\phi_{0}$ in the absolute lab-frame coordinates. Thus, the ionic potential 
not only induces a focusing of the indirect electrons, considerably narrowing their momentum distribution, but also an 
angular streaking of the distribution. Does the interplay of focusing and angular streaking dynamics, which are both
due to the ionic potential, explain the large asymmetry observed in the Coulomb-focused signal induced 
by the OTC fields? 

The attoclock streaking is determined by the scattering of the outgoing electrons in the ionic potential. 
In chiral molecules, this scattering induces an asymmetry between the number of electrons ending up in the 
forward and backward hemispheres -- the well known PECD effect. 
It was recently shown that the attoclock streaking angle is also forward/backward asymmetric 
\cite{fehre2019b,bloch2021}. The chirality of the potential causes a shift $\delta\phi_{chiral}$ of the forward 
electrons, and an opposite shift of the backward electrons. This is schematized in Fig. \ref{figstreak}(b,c), 
which presents the result of a rotation by $\pm 4^\circ$ of the momentum distribution from Fig. \ref{figstreak}(a). Note that this figure does not present the results of actual photoionization calculations, but aims at pedagogically describing the underlying process. 
The chiral attoclock shift produces a  $\Delta P^{f/b}$ asymmetry, obtained by subtracting the forward and backward 
distributions (Fig \ref{figstreak}(d)). Note that this asymmetry appears even if the total number of electrons 
ejected forward and backward is the same. The $\delta\phi_{chiral}$ shift, being a chirosensitive quantity, it 
reverses when switching from one enantiomer to its mirror image, or when inverting the helicity of the ionizing 
radiation. However, in the latter case, one has to keep in mind that this reversal is relative to the laser 
field rotation direction. This can be intuitively understood by considering the angular shift as a delay. In the 
schematic representation of Fig. \ref{figstreak}(b-c), the forward electrons appear at later angles along 
the rotation of the vector potential. By analogy with a clock, these electrons are positively delayed. 
The situation must reverse when switching the light helicity \cite{fehre2019b,bloch2021}. The forward electrons 
must appear at earlier angles along the rotation of the vector potential. As a consequence, the chiral angular 
shift $\delta\phi_{chiral}$ does not change sign in the absolute coordinate system when reversing the light 
helicity (see Figs. 3(f-g)). Comparing the differential distributions $\Delta P^{f/b}$ displayed in Figs. 3(d,h) 
then shows that with both helicities, the forward electrons are deviated in the counter-clockwise direction with 
respect to the backward electrons. This shift would change sign upon switching of the enantiomer. Indeed, while 
switching the light or molecular handedness is equivalent in chiral photoionization by circularly polarized radiation, 
these two operations have different effects in elliptical (or more complex) fields \cite{comby2018a}. 
Resolving the forward/backward asymmetry in the ($p_x,p_y$)-plane thus produces a chirosensitive signal which 
enables the determination of the absolute configuration of the enantiomer, independently of the helicity of the 
ionizing radiation. 

How should we expect the chiral attoclock shift to play a role in the OTC field case? The streaking dynamics
are displayed in Figs. (i-l) for $\varphi_{\omega/2\omega}=\pi$. In the upper hemisphere, 
the situation is similar to the one described in Fig.\ref{figstreak}(a-d). The vector potential rotates 
counter-clockwise and we assume that a $\pm\delta\phi_{chiral}$ shift is induced between electrons ejected 
forward and backward by molecular chirality. In the lower hemisphere, the vector potential rotates clockwise. 
As we have just seen, whereas the non-chiral part of the streaking angle reverses between the upper and lower 
hemispheres, the chiral angular shift goes in the same direction. Chirality induces a rotation of the whole 
angular momentum distribution (Fig. \ref{figstreak}(j,k)), just like in the single helicity case of 
Figs. \ref{figstreak}(b,c). As a consequence, the differential distribution $\Delta P^{f/b}$ shows four main 
quadrants, whose sign depends on the absolute configuration of the enantiomer. Adding $\pi$ to the selected
two-color phase reverses the OTC field rotation. As shown in Figs. 3(m-p), $\phi_0$ is accordingly changed to 
$-\phi_0$ in both hemispheres while $\delta\phi_{chiral}$ remains unchanged. The streaking of forward electrons
in the $\varphi_{\omega/2\omega}+\pi$ case is then symmetric to the one of backward electrons for 
$\varphi_{\omega/2\omega}$ with respect to the $y$-axis, and vice-versa. We thus understand that when changing
$\varphi_{\omega/2\omega}$ to $\varphi_{\omega/2\omega}+\pi$, the interplay of isotropic ($\phi_0$) and chiral 
($\delta\phi_{chiral}$) streakings leads to asymmetric distributions which are negative mirror-images of each other
with respect to the Coulomb focusing direction.

\begin{figure}[t]\centering 
	\includegraphics[width=0.8\linewidth]{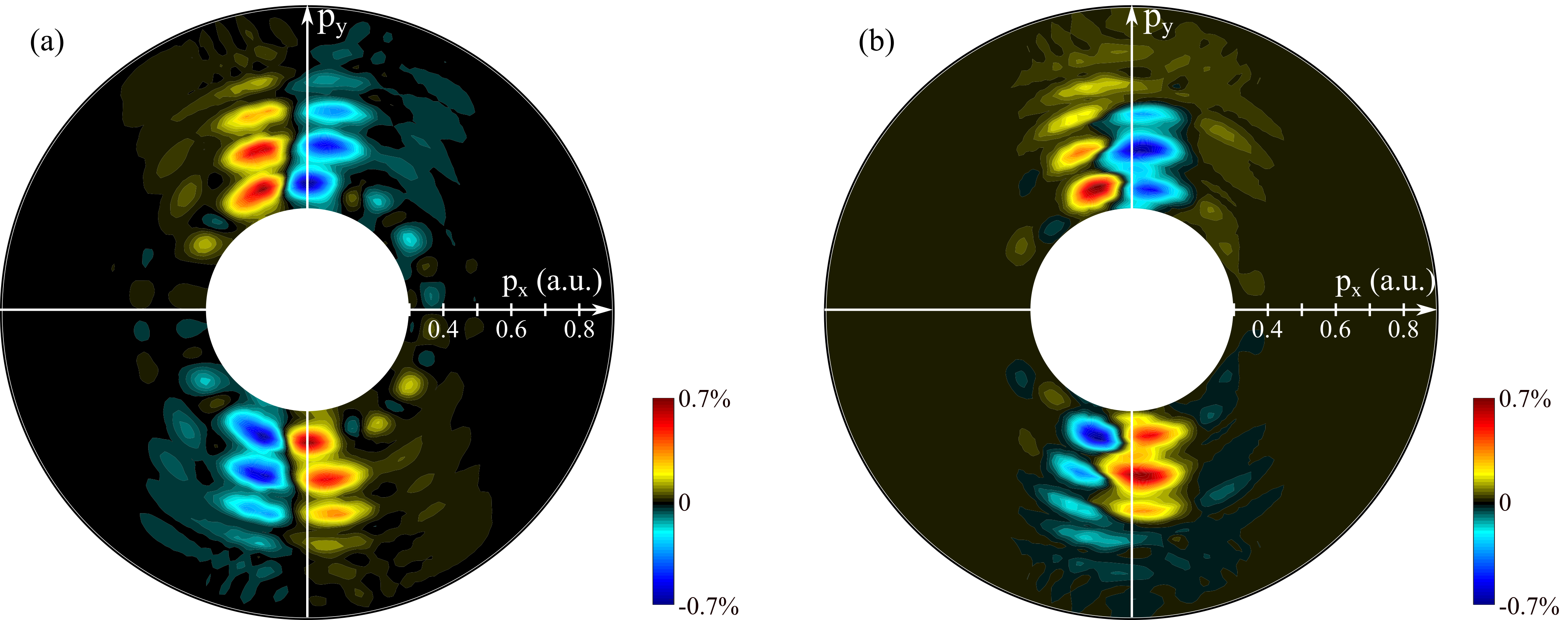}
	\caption{Forward/backward asymmetry of the electron momentum distribution, obtained from the simple model assuming a chiral-induced rotation of the distribution (a) and from full TDSE calculations with an OTC field at 5$\times 10^{13}$ W.cm$^{-2}$, $r=0.1$ and $\varphi_{\omega/2\omega}=\pi$. The central part of the distributions is not shown to draw attention to the momentum range where Coulomb focusing operates on indirect electrons.}
	\label{figcompare}
\end{figure}

The analysis presented in Fig. \ref{figstreak} is qualitative. It assumes that the only effect of molecular 
chirality is to induce a forward/backward asymmetric rotation of the photoelectron angular momentum distribution. 
This description presents a simplistic picture of the interaction. In general, the number of electrons ejected 
forward and backward differs -- this is the essence of PECD. Furthermore, the interference between direct and 
indirect electrons can play an important role in the chiroptical signal. Last, all these effects generally 
strongly depend on the kinetic energy of the ejected electrons. In spite of these multiple effects, the 
differential distribution obtained in Fig. \ref{figstreak}(l) is remarkably similar to the actual result of 
the TDSE calculation, as can be seen in Fig. \ref{figcompare}. The peak value of $\Delta P^{f/b}$ can be adjusted 
in the model by setting the value of the $\delta\phi_{chiral}$ rotation. A good agreement is obtained for 
$\delta\phi_{chiral}=0.07^\circ$. The chiral signal is almost exclusively located along the Coulomb-focused component. 
The model predicts some chiral signal along the interference fringes in the first ATI peak , but they are absent 
from the TDSE calculation (see Fig. 1). This strengthens the conclusion that the chirality of the potential mostly affects 
the Coulomb-focused indirect electrons, which are too localized to produce significant interference patterns. 

These results demonstrate that the indirect electrons, which are Coulomb-focused, are angularly streaked by the 
rotating electric field, and that this process is sensitive to the chirality of the potential. The electrons 
ending up forward are angularly shifted by $0.14^\circ$ with respect to the electrons ending up backwards. 
This small angular shift is sufficient to induce a normalized forward/backward asymmetry 
$\Delta P^{f/b}$ in the $1\%$ range, which is the typical order of magnitude of a strong-field PECD signal. 

\section{Conclusion}
In this study we have resolved the structural dependence of the Coulomb focusing phenomenon that plays an
important role in a large range of strong-field interactions. 
Drawing from the work of Richter \textit{et al.} who showed that OTC strong fields are particularly adequate 
to the observation of Coulomb-focused indirect electrons \cite{richter2016}, we have numerically investigated 
the sensitivity of such electrons to the chirality of a toy-model molecule. We have found that these electrons 
present an enhanced sensitivity to chirality, which is almost independent to the two-color phase of the OTC field. 
We traced back the root of this sensitivity to a forward/backward asymmetric angular streaking, induced by the 
laser-assisted scattering of the electrons onto the chiral ionic potential. 

We ascertained from a quantum mechanical point of view that Coulomb focusing operates at long range, out of 
the molecular core. On the other hand, large chiral features are expected to be imprinted in the photoelectron signal
at short range. However one must refrain from interpreting our results in terms of a two-step process 
composed of focusing by the isotropic long-range part of the potential, and chiral response at the vicinity of 
the ionic core.
Chiral patterns survive in the ionic 
potential far away from the core. Even if they decrease faster than the isotropic term as the radial distance $r$ 
increases, these asymmetric components influence continuously the electron motion. 
Coulomb focusing and chiral
scattering are thus intertwined, and the returning electrons are focused by the whole electrostatic force exerted
by the ion core, including its isotropic (Coulombic) and anisotropic (eventually chiral) components.

We have shown that electrons experiencing laser-assisted elastic scattering onto a chiral ionic core
exhibit a large chiroptical response. This could be demonstrated from an experimental point of view, using OTC driving fields and 3D photoelectron momentum imaging as in \cite{richter2016}. More generally, one can expect Coulomb focusing to play a role in the chiral signal whenever the strong-field radiation is not circular. For instance, it could influence the electron dynamics driven by locally and globally chiral fields, which provide gigantic chiral responses in high-harmonic generation \cite{ayuso2019}.

In this work we focused on electrons with rather low asymptotic energies, belonging to
the first ATI peaks. Future works should investigate the chiral response of higher energy electrons,
including those who backscatter onto the ionic core and present asymptotic energies up to $\sim 10 U_P$, where
$U_P$ is the ponderomotive energy of the driving pulse \cite{paulus1994}. Among these electrons, some of them
have de Broglie wavelengths appropriate to self-image the target structure by diffraction. This would enlarge
the scope of laser-induced electron diffraction \cite{meckel2008,blaga2012} to chiral imaging.

\section{Acknowledgements}
We acknowledge S. Beaulieu, V. Blanchet and E. Bloch for fruitful discussion. This project has received funding from the European Research Council (ERC) under the European Union's Horizon 2020 research and innovation program no. 682978 - EXCITERS, 864127 - ATTOGRAM and from 871124 - Laserlab-Europe. N. D. is the incumbent of the Robin Chemers Neustein Professorial Chair. N. D. acknowledges the Minerva Foundation, the Israeli Science Foundation.

\bibliography{BetterBibtexFromZotero}

\end{document}